\documentclass[aps,prb,twocolumn,reprint,superscriptaddress]{revtex4-2}

\usepackage{graphicx}
\usepackage{amsmath,amssymb}
\usepackage{textcomp}          
\usepackage{color}             

\makeatletter
\def\fig@none{@@none@@}
\let\figfound\fig@none
\newcommand*\FIGtry[1]{%
  \ifx\figfound\fig@none
    \IfFileExists{#1.pdf}{\xdef\figfound{#1}}{%
    \IfFileExists{#1.png}{\xdef\figfound{#1}}{%
    \IfFileExists{#1.jpg}{\xdef\figfound{#1}}{}}}%
  \fi}
\newcommand*\FIG[2]{%
  \global\let\figfound\fig@none
  \FIGtry{figures/fig#1}\FIGtry{figures/#1}\FIGtry{fig#1}\FIGtry{#1}%
  \ifx\figfound\fig@none
    \PackageWarning{28Si-manuscript}{Fig. #1: image file not found, placeholder inserted}%
    \fbox{\parbox{0.9\linewidth}{\centering\ttfamily\footnotesize
      Fig. #1 --- image file not found.\newline
      Tried \texttt{figures/fig#1}, \texttt{figures/#1}, \texttt{fig#1},
      \texttt{#1} with .pdf/.png/.jpg}}%
  \else
    \includegraphics[#2]{\figfound}%
  \fi}
\makeatother

\begin{document}

\title{High-purity \textsuperscript{28}Si epilayers (\textsuperscript{29}Si $\leq$ 30 ppm) by industrial RPCVD for high-fidelity electron spin qubits}

\author{Zhenzhen Kong}
\affiliation{Beijing Superstring Academy of Memory Technology, Beijing 100176, P. R. China}
\affiliation{Hefei National Laboratory, Hefei 230088, P. R. China}
\author{Weizhu Liao}
\affiliation{Laboratory of Quantum Information, University of Science and Technology of China, Hefei 230026, P. R. China}
\author{Ning Chu}
\affiliation{Laboratory of Quantum Information, University of Science and Technology of China, Hefei 230026, P. R. China}
\author{Yiwen Zhang}
\affiliation{Beijing Superstring Academy of Memory Technology, Beijing 100176, P. R. China}
\affiliation{Hefei National Laboratory, Hefei 230088, P. R. China}
\author{Ying Zhang}
\affiliation{Beijing Superstring Academy of Memory Technology, Beijing 100176, P. R. China}
\author{Jixuan Li}
\affiliation{Beijing Superstring Academy of Memory Technology, Beijing 100176, P. R. China}
\affiliation{Laboratory of Quantum Information, University of Science and Technology of China, Hefei 230026, P. R. China}
\author{Changyang Xu}
\affiliation{Beijing Superstring Academy of Memory Technology, Beijing 100176, P. R. China}
\affiliation{Laboratory of Quantum Information, University of Science and Technology of China, Hefei 230026, P. R. China}
\author{Yongliang Hao}
\affiliation{Beijing Superstring Academy of Memory Technology, Beijing 100176, P. R. China}
\affiliation{Integrated Circuit Advanced Process R\&D Center, Institute of Microelectronics, Chinese Academy of Sciences, Beijing 100029, P. R. China}
\author{Yi Zhou}
\affiliation{Beijing Superstring Academy of Memory Technology, Beijing 100176, P. R. China}
\affiliation{Laboratory of Quantum Information, University of Science and Technology of China, Hefei 230026, P. R. China}
\author{Yanpeng Song}
\affiliation{Beijing Superstring Academy of Memory Technology, Beijing 100176, P. R. China}
\author{Xiaomeng Liu}
\affiliation{Beijing Superstring Academy of Memory Technology, Beijing 100176, P. R. China}
\author{Hailing Wang}
\affiliation{Beijing Superstring Academy of Memory Technology, Beijing 100176, P. R. China}
\author{Xiangsheng Wang}
\affiliation{Beijing Superstring Academy of Memory Technology, Beijing 100176, P. R. China}
\author{Jinbiao Liu}
\affiliation{Laboratory of Quantum Information, University of Science and Technology of China, Hefei 230026, P. R. China}
\author{Chao Zhao}
\affiliation{Beijing Superstring Academy of Memory Technology, Beijing 100176, P. R. China}
\author{Hao Meng}
\affiliation{Beijing Superstring Academy of Memory Technology, Beijing 100176, P. R. China}
\author{Gang Cao}
\affiliation{Hefei National Laboratory, Hefei 230088, P. R. China}
\affiliation{Laboratory of Quantum Information, University of Science and Technology of China, Hefei 230026, P. R. China}
\author{Guo-Ping Guo}
\affiliation{Hefei National Laboratory, Hefei 230088, P. R. China}
\affiliation{Laboratory of Quantum Information, University of Science and Technology of China, Hefei 230026, P. R. China}
\affiliation{Origin Quantum Computing Company Limited, Hefei 230088, P. R. China}
\author{Hai-Ou Li}
\thanks{Corresponding authors. E-mail: wangguilei@hfnl.cn (G. Wang); haiouli@ustc.edu.cn (H.-O. Li).}
\affiliation{Hefei National Laboratory, Hefei 230088, P. R. China}
\affiliation{Laboratory of Quantum Information, University of Science and Technology of China, Hefei 230026, P. R. China}
\author{Guilei Wang}
\thanks{Corresponding author.}
\affiliation{Beijing Superstring Academy of Memory Technology, Beijing 100176, P. R. China}
\affiliation{Hefei National Laboratory, Hefei 230088, P. R. China}

\date{\today}

\begin{abstract}
Silicon quantum-dot spin qubits are among the most promising routes to large-scale quantum computation, but the residual \textsuperscript{29}Si nuclear spins in natural silicon impose a hyperfine dephasing limit that must be lifted by isotopic purification. Here we report the epitaxial growth of high-purity \textsuperscript{28}Si on float-zone silicon using an industrial reduced-pressure chemical vapour deposition (RPCVD) system. Secondary-ion mass spectrometry (SIMS) confirms an epilayer \textsuperscript{28}Si isotopic purity of 99.997\%, with \textsuperscript{29}Si suppressed below 30 ppm. Through a systematic vacuum annealing matrix (850--1150 \textdegree{}C), combined with atomic-force microscopy and SIMS, we map the defect-mediated decomposition of the native oxide and the injection of non-equilibrium vacancies, and we identify a thermal-budget window below 950 \textdegree{}C within which the thin-layered structure of \textsuperscript{28}Si remains intact. At $\sim$1.5 K, magneto-transport reveals a peak Hall mobility of $\sim$12,858 cm\textsuperscript{2}V\textsuperscript{-1}s\textsuperscript{-1} and a quantum scattering lifetime of 0.659 ps (single-particle level broadening 499 \textmu{}eV). A double quantum dot fabricated on this platform exhibits T\textsubscript{2}* = 0.62 \textmu{}s, a Hahn-echo coherence time T\textsubscript{2}\textsuperscript{Hahn} = 28.8 \textmu{}s, and randomized-benchmarking single-qubit gate fidelities up to 99.9\%. Our results establish industrial RPCVD as a promising route toward high-coherence, high-fidelity silicon spin qubits.
\end{abstract}

\maketitle


\section{Introduction}

Quantum computation offers a decisive advantage for the simulation of specific complex systems and for certain accelerated algorithms. Among the hardware platforms under active investigation, CMOS-compatible semiconductor quantum-dot qubits are widely regarded as one of the most credible solid-state architectures for large-scale monolithic integration, owing to their exceptional scalability [\textcolor{blue}{1--3}]. Realizing high-fidelity qubit control, however, depends critically on an ultra-low-disorder electromagnetic and lattice environment.

In natural silicon, the non-zero-nuclear-spin isotope \textsuperscript{29}Si (nuclear spin I = 1/2, natural abundance $\approx$ 4.673\%) [\textcolor{blue}{4,5}] is spatially disordered and therefore generates a randomly fluctuating magnetic field acting on the localized electron. For an electron strongly confined in a quantum dot, this fluctuating nuclear-spin bath constitutes the dominant hyperfine relaxation channel and is one of the core mechanisms limiting qubit coherence (T\textsubscript{2}* $\approx$ 1 \textmu{}s in natural silicon) [\textcolor{blue}{6}]. Removing \textsuperscript{29}Si by isotopic purification [\textcolor{blue}{4,39}] --- thereby suppressing the random nuclear-field perturbation --- is the most effective and widely adopted strategy for extending qubit dephasing times, and can improve coherence times by one to two orders of magnitude [\textcolor{blue}{7--9}].

To satisfy the large-scale requirements of universal quantum computing, the preparation of \textsuperscript{28}Si must be transitioned from laboratory-scale, specialized equipment to high-throughput, standardized, industrial semiconductor manufacturing platforms [\textcolor{blue}{10,11}]. Here, we report the epitaxial growth of an high-purity \textsuperscript{28}Si-MOS structure using an industrial reduced-pressure chemical vapor deposition (RPCVD) system, obtaining epilayers with an isotopic purity as high as 99.997\%. To establish the thermal-budget limits of this system during subsequent device processing and to clarify the microscopic structural evolution, we designed and executed a systematic vacuum annealing matrix. By combining real-space and reciprocal-space characterization, we traced the coupling between thermal decomposition of the native surface oxide, non-equilibrium vacancy injection, and the physical boundaries of local defect nucleation, long-range isotopic interdiffusion, and dewetting at extreme temperatures. In addition, low-temperature (down to 10 mK) magnetotransport was used to study the low-temperature mobility and deep microscopic scattering mechanisms of an induced two-dimensional electron gas (2DEG) at electron inversion layer forms near the \textsuperscript{28}Si/\textsuperscript{28}SiO\textsubscript{2} interface, confirming an low disorder level in the channel. Finally, we fabricated lateral double-quantum-dot devices on the optimized material platform, extracted the electron-spin dephasing and coherence times, and quantitatively evaluated single-qubit gate fidelity by standard randomized benchmarking (RB). These results show that industrial RPCVD-grown \textsuperscript{28}Si films constitute a promising material platform for building high-fidelity spin qubits [\textcolor{blue}{13,22}].

\section{Results}

\subsection{Epitaxial growth and structural characterization}

Epitaxial growth was performed in an industrial RPCVD system using high-resistivity float-zone silicon (FZ-Si, resistivity $>$ 10,000 $\Omega$\textperiodcentered{}cm) as the substrate. The isotopically enriched layer used high-purity \textsuperscript{28}SiH\textsubscript{4} as the precursor, which was converted from gas-centrifuge-enriched SiF\textsubscript{4} feedstock [\textcolor{blue}{4,10}]. Mass-spectrometry analysis confirmed that the \textsuperscript{28}Si isotopic abundance of the precursor reached 99.99772\% $\pm$ 0.00025\%, while the residual \textsuperscript{29}Si and \textsuperscript{30}Si abundances were suppressed to 0.00221\% $\pm$ 0.00025\% and 0.00007\% $\pm$ 0.00005\%, respectively.

\begin{figure*}[t]
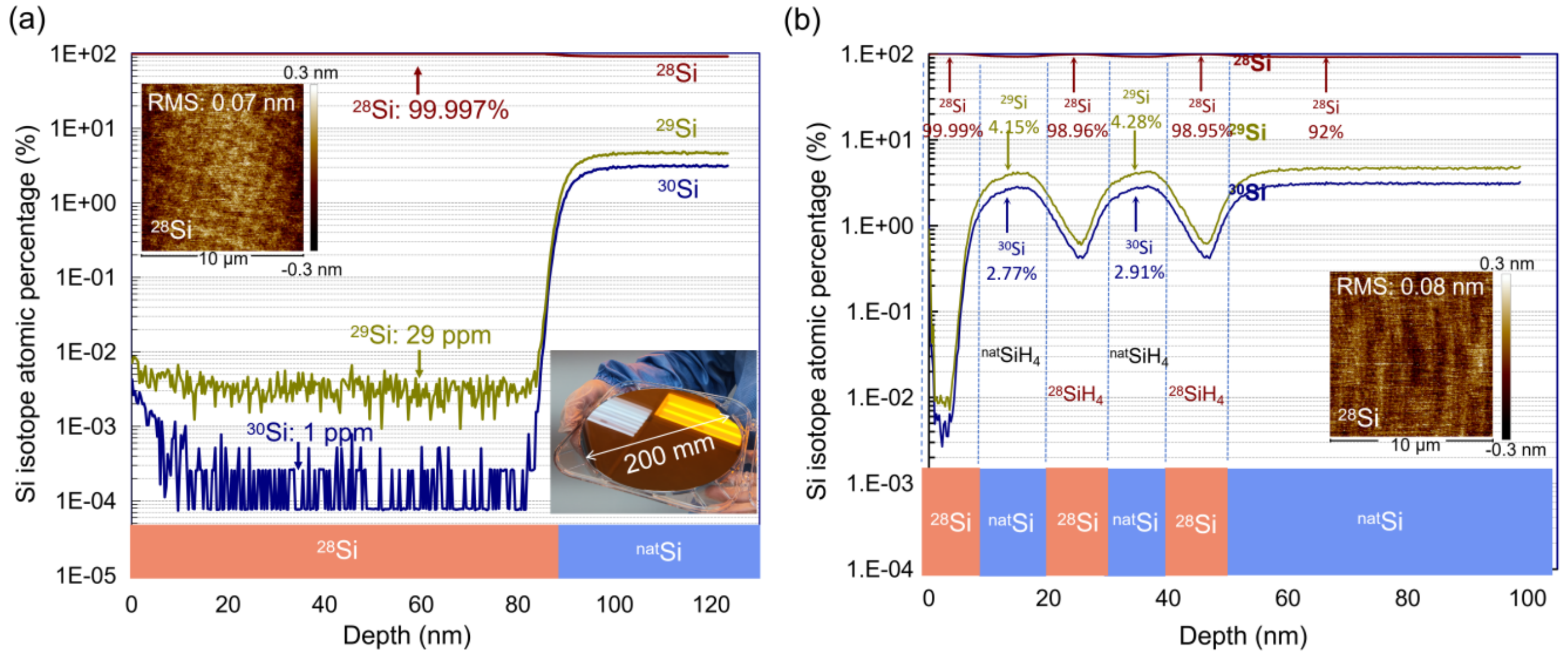

  \centering
  \FIG{1}{width=\textwidth}
  \caption{\textbf{Isotopic and morphological characterization of the enriched \textsuperscript{28}Si single epilayer and the \textsuperscript{28}Si/natSi isotope superlattice.} (a) SIMS depth profile of the highly enriched \textsuperscript{28}Si single epilayer grown on a natural-silicon (natSi) substrate. The \textsuperscript{28}Si purity in the epilayer reaches 99.997\%, with residual \textsuperscript{29}Si and \textsuperscript{30}Si concentrations suppressed to 29 ppm and 1 ppm, respectively. The bottom schematic depicts the layered stack; the left inset shows an AFM image of the epilayer surface with an atomically smooth morphology (RMS = 0.07 nm); the right inset shows a real photo of the purified \textsuperscript{28}Si isotope epitaxial layer grown on an 8-inch FZ Si substrate using industrialized RPCVD technology. (b) SIMS depth profile of the \textsuperscript{28}Si/\textsuperscript{nat}Si isotope superlattice grown by alternating \textsuperscript{28}SiH\textsubscript{4} and natSiH\textsubscript{4} precursors. The vertical dividers and bottom schematic delineate the periodic modulation of the isotope concentrations (\textsuperscript{28}Si, \textsuperscript{29}Si and \textsuperscript{30}Si) and the sharp interface profiles between layers. The inset shows the AFM surface morphology (RMS = 0.08 nm).}
  \label{1}
\end{figure*}

Two microstructures grown at 650 \textdegree{}C were designed to investigate the intrinsic atomic self-diffusion and vacancy-migration mechanisms (Fig.~\ref{1}, bottom schematics): (i) a \textbf{single homoepitaxial layer} consisting of an $\sim$83 nm thick highly enriched \textsuperscript{28}Si epilayer, and (ii) an \textbf{isotope superlattice} constructed by alternating the \textsuperscript{28}SiH\textsubscript{4} and \textsuperscript{nat}SiH\textsubscript{4} precursors to build a multilayer heterostructure composed of three 10 nm \textsuperscript{28}Si layers separated by two 10 nm natural silicon (\textsuperscript{nat}Si) barrier layers. The multilayer was subsequently annealed under high vacuum to probe its mass-transport and defect dynamics.

SIMS depth profiling (Fig.~\ref{1}(a)) shows a highly enriched \textsuperscript{28}Si epilayer with a stable isotopic ratio of 99.997\% extending from the surface to $\sim$83 nm. Within this ultra-pure region, the impurity isotopes \textsuperscript{29}Si and \textsuperscript{30}Si are suppressed to $\sim$29 ppm and $\sim$1 ppm, respectively (nominal values calibrated against standard samples; the relative uncertainty increases near the SIMS detection limit). A steep isotopic transition is observed between 83 nm and 95 nm, across which the \textsuperscript{29}Si and \textsuperscript{30}Si concentrations rise by several orders of magnitude before stabilizing at their natural abundances in the underlying \textsuperscript{nat}Si substrate (\textsuperscript{29}Si $\approx$ 4.7\%, \textsuperscript{30}Si $\approx$ 3.1\%). This sharp concentration gradient reflects a well-defined, diffusion-limited interface formed during growth. The atomic-force-microscopy (AFM) inset shows an atomically smooth surface with a root-mean-square (RMS) roughness of only 0.07 nm.

Figure~\ref{1}(b) presents a quantitative evaluation of the isotopic composition and depth distribution of the \textsuperscript{28}Si/\textsuperscript{nat}Si multilayer. The \textsuperscript{29}Si and \textsuperscript{30}Si atomic percentages oscillate strongly and periodically, in exact agreement with the designed multilayer stack. The topmost enriched \textsuperscript{28}Si layer (0--10 nm) reaches an isotopic purity as high as 99.99\%, while the embedded quantum-well-like \textsuperscript{28}Si layers (at $\sim$25 nm and $\sim$47 nm) reach 98.96\% and 98.95\%, respectively. The interleaved natSi barrier layers show peak \textsuperscript{29}Si concentrations of 4.15\%--4.28\% and \textsuperscript{30}Si of 2.77\%--2.91\%. The slight smoothing of the concentration peaks and valleys at the nanometre-scale interfaces is attributed mainly to instrumental cascade mixing during SIMS sputtering and to weak thermal interdiffusion; nevertheless, the high-contrast compositional modulation confirms atomically abrupt interfaces and excellent growth-control precision [\textcolor{blue}{12}]. The inset shows that the surface quality is equally well preserved (RMS = 0.08 nm).

\begin{figure*}[t]
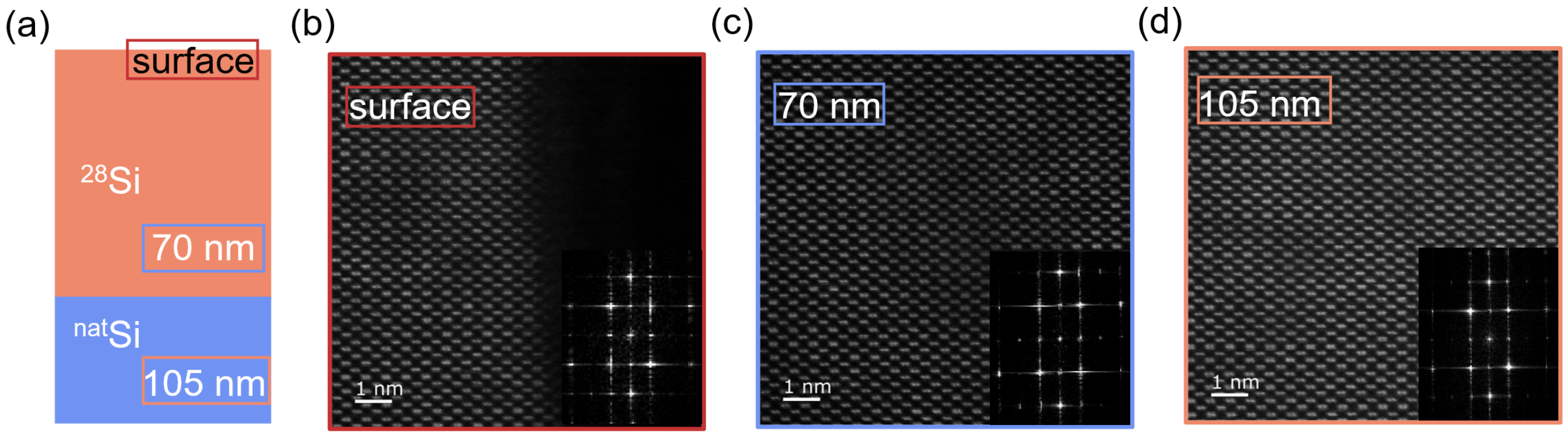

  \centering
  \FIG{2}{width=\textwidth}
  \caption{\textbf{Cross-sectional microstructure and crystallographic analysis of the enriched \textsuperscript{28}Si/\textsuperscript{nat}Si epitaxial heterostructure.} (a) Schematic cross-section indicating the specific regions and depths (surface, 70 nm and 105 nm) selected for high-resolution TEM characterization. (b--d) Atomically resolved HRTEM images taken at (b) the film surface, (c) 70 nm depth within the enriched \textsuperscript{28}Si epilayer, and (d) 105 nm depth within the \textsuperscript{nat}Si substrate. The atom columns aligned along the [110] zone axis reveal a highly ordered, defect-free diamond-cubic lattice. Insets: corresponding fast-Fourier-transform (FFT) patterns with symmetric, sharp diffraction spots confirming the exceptional crystalline perfection and coherent lattice-matched epitaxial relationship of the entire heterostructure.}
  \label{2}
\end{figure*}

To evaluate the crystalline quality and lattice coherence of the epitaxial structures, cross-sectional high-resolution transmission electron microscopy (HRTEM) with fast-Fourier-transform (FFT) analysis was performed at three representative depths --- the film surface (Fig.~\ref{2}(b)), 70 nm inside the epilayer (Fig.~\ref{2}(c)), and 105 nm inside the \textsuperscript{nat}Si substrate (Fig.~\ref{2}(d)). Atomically resolved lattice images along the [110] zone axis show a highly ordered, defect-free diamond-cubic lattice in all regions. Critically, the corresponding FFT patterns (insets, Fig.~\ref{2}(b--d)) exhibit sharp, symmetric diffraction spots with no detectable spot splitting, distortion, or phase separation across the entire heterostructure. This near-perfect lattice continuity demonstrates that the isotopic modulation introduces no coherent strain or structural dislocations, confirming that the \textsuperscript{28}Si/\textsuperscript{nat}Si superlattice possesses exceptional crystalline perfection. This high-quality, strain-free heterostructure thus provides an ideal model system in which extrinsic-defect-assisted diffusion pathways are eliminated, laying a solid foundation for the study of intrinsic atomic self-diffusion and vacancy-migration mechanisms in the subsequent annealing experiments.

\subsection{Thermal stability and interdiffusion}

\begin{figure*}[t]
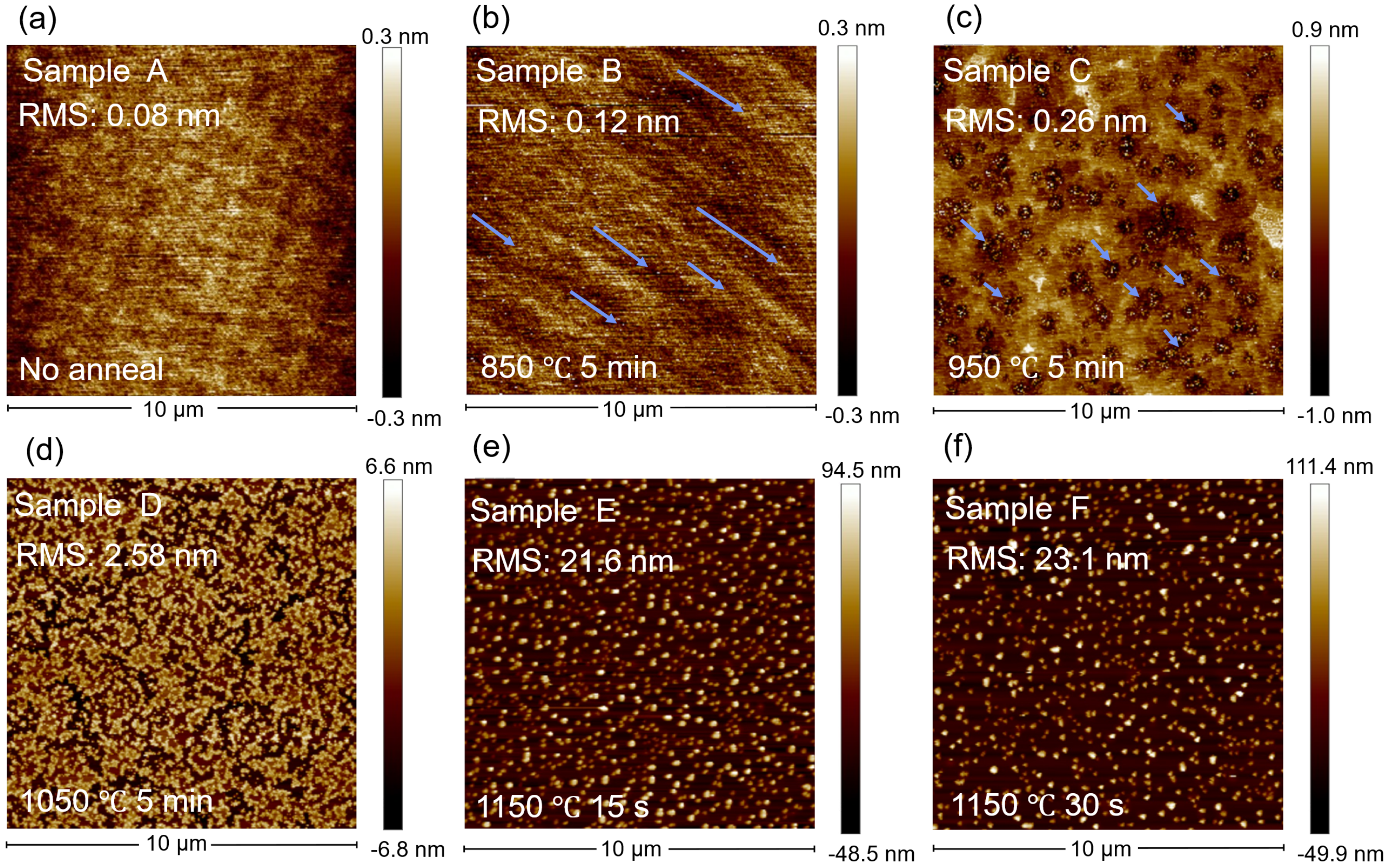

  \centering
  \FIG{3}{width=\textwidth}
  \caption{\textbf{Thermal evolution of the surface morphology and morphological stability of the \textsuperscript{28}Si/\textsuperscript{nat}Si multilayer under vacuum annealing.} (a) As-grown reference (Sample A), atomically smooth with RMS = 0.08 nm. (b) Annealed at 850 \textdegree{}C for 5 min (Sample B, RMS = 0.12 nm); blue arrows mark the periodic atomic steps and terraces formed by surface relaxation. (c) Annealed at 950 \textdegree{}C for 5 min (Sample C, RMS = 0.26 nm); blue arrows highlight the nucleation of nanoscale surface pits driven by vacancy aggregation. (d) Annealed at 1050 \textdegree{}C for 5 min (Sample D, RMS = 2.58 nm), showing a transition to a coarsened, interconnected maze-like network. (e,f) Annealed at the extreme temperature of 1150 \textdegree{}C for (e) 15 s (RMS = 21.6 nm) and (f) 30 s (RMS = 23.1 nm), showing severe agglomeration and disintegration into isolated 3D islands by dewetting.}
  \label{3}
\end{figure*}

To study the structural stability and surface chemistry of the isotopically engineered \textsuperscript{28}Si/\textsuperscript{nat}Si multilayer, a systematic vacuum annealing matrix was designed and executed. The as-grown structure naturally carries a thin silicon dioxide (SiO\textsubscript{x}) native oxide on its outermost surface; this native oxide is only a few nanometres thick, so that under high vacuum it decomposes at elevated temperature and leaves the voids observed by AFM. By varying the annealing temperature from 850 \textdegree{}C to 1150 \textdegree{}C and modulating the annealing time, the experiment strategically captures the real-space kinetic evolution from early local nucleation of oxide-decomposition voids to the morphological collapse of the underlying silicon film. The annealed surface morphology was quantitatively characterized by AFM and is summarized in Fig.~\ref{3}(a--f).

The as-grown reference sample (Fig.~\ref{3}(a)) exhibits a near-perfect, atomically smooth surface (RMS = 0.08 nm). After annealing at 850 \textdegree{}C for 5 min (Fig.~\ref{3}(b)), the surface remains smooth (RMS = 0.12 nm) and clearly develops parallel atomic steps and terraces (blue arrows), driven by initial local surface relaxation. Upon increasing the temperature to 950 \textdegree{}C (Fig.~\ref{3}(c)), the RMS roughness rises slightly to 0.26 nm and numerous nanoscale dark pits nucleate on the surface (blue arrows). Unlike the laterally uniform silicon oxidation, decomposition of the SiO\textsubscript{x} layer is highly locally non-uniform: pre-existing intrinsic defects within the native oxide or at the SiO\textsubscript{x}/Si interface (micro-pores or trace metal contamination) act as central nucleation sites for the decomposition reaction. At these defect centres, the interfacial reaction Si + SiO\textsubscript{x} $\to$ 2SiO$\uparrow$ is activated; the resulting silicon monoxide (SiO) is highly volatile and rapidly sublimes into the vacuum, leaving localized nanoscale voids at the interface that appear as nucleated dark pits in the AFM images. The injected non-equilibrium vacancies may additionally be captured by interstitial oxygen to form stable vacancy--oxygen (Si-A) centres [\textcolor{blue}{15}].

At 1050 \textdegree{}C (Fig.~\ref{3}(d)), the AFM morphology reveals a structural instability, with the surface transforming into a coarsened, interconnected maze-like network and the RMS roughness jumping to 2.58 nm. This morphological transition represents the lateral expansion and coalescence of the oxide-decomposition voids. single-crystal silicon atoms acquire liquid-like surface self-diffusivity at such high temperature; driven by chemical-potential gradients, the exposed silicon atoms migrate rapidly outward from the pore edges, accelerating the lateral rupture of the continuous film. At the extreme temperature of 1150 \textdegree{}C (Fig.~\ref{3}(e,f)), the native oxide is completely removed within seconds by total SiO volatilization, and the fully exposed, unconfined ultra-thin silicon epilayer disintegrates under capillary forces and a Rayleigh-like capillary instability [\textcolor{blue}{18}]. Driven by the liquid-like mobility of silicon adatoms, the film undergoes dewetting, agglomerating within 15 s into isolated three-dimensional (3D) hemispherical islands (RMS = 21.6 nm) that stabilize at 30 s (RMS = 23.1 nm). This evolution path therefore validates the classical defect-mediated oxide-decomposition model (Tromp et al., 1985) [\textcolor{blue}{14}]. To ensure that the layered geometry reflects pure, solid-state, intrinsic bulk diffusion --- rather than being contaminated by defect-triggered oxide rupture and surface mass-transport artefacts.

\begin{figure}[t]
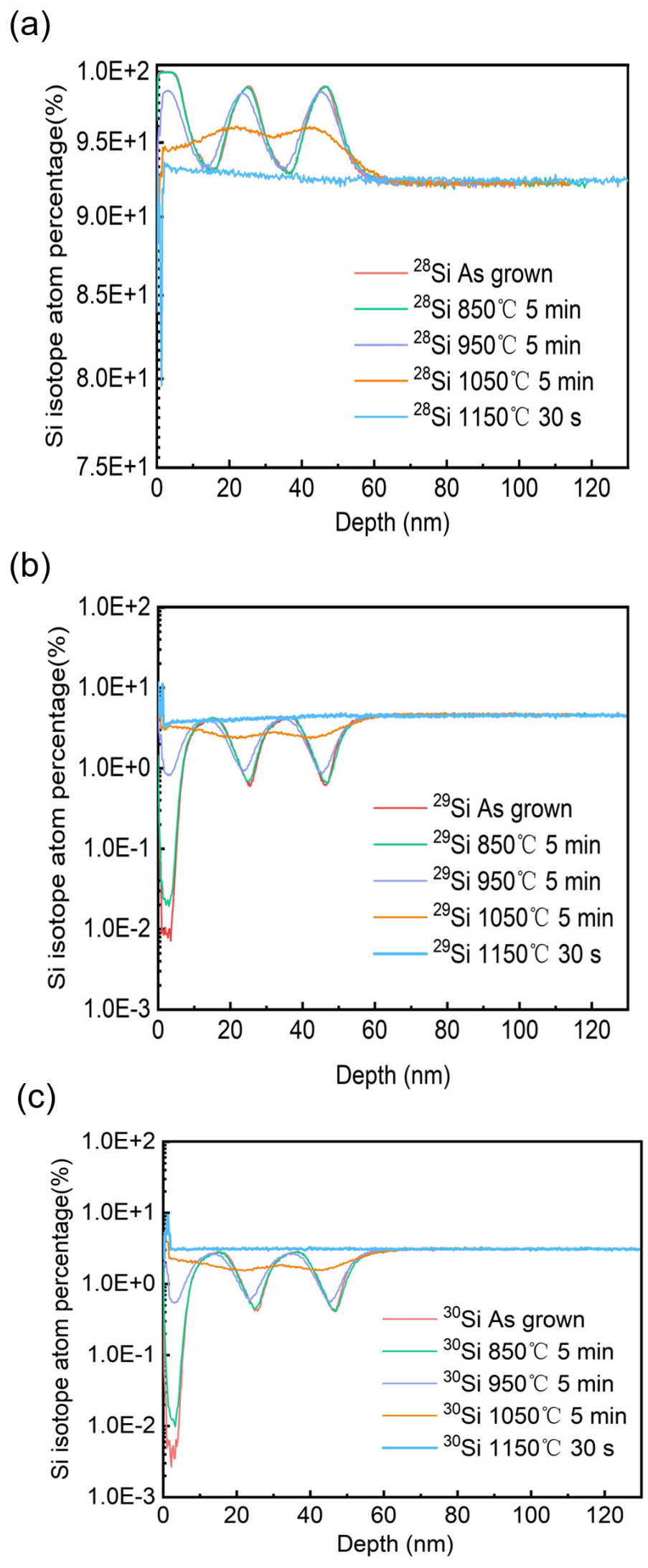

  \centering
  \FIG{4}{height=0.62\textheight,keepaspectratio}
  \caption{\textbf{SIMS depth profiles showing the isotopic-composition evolution of the \textsuperscript{28}Si/\textsuperscript{nat}Si isotope superlattice under different thermal budgets.} (a--c) Depth profiles of (a) \textsuperscript{28}Si , (b) \textsuperscript{29}Si and (c) \textsuperscript{30}Si for the as-grown reference and after annealing at 850--1150 \textdegree{}C. The evolution from clear square-wave periodic oscillations to a fully homogenized horizontal line directly reflects the transition from local interface broadening to macroscopic structural mixing.}
  \label{4}
\end{figure}

To quantitatively track the spatial mass transport and redistribution of silicon isotopes across the designed interfaces, SIMS depth profiling was systematically performed on the \textsuperscript{28}Si/\textsuperscript{nat}Si isotope superlattice. As shown in Fig.~\ref{4}(a--c), the as-grown reference sample and the 850 \textdegree{}C-annealed film exhibit excellent periodic square-wave oscillations in the \textsuperscript{28}Si, \textsuperscript{29}Si and \textsuperscript{30}Si depth profiles over the outermost $\sim$50 nm. The \textsuperscript{28}Si concentration reaches near 100\% in the enriched regions, while the minor isotopes \textsuperscript{29}Si and \textsuperscript{30}Si are deeply suppressed to background levels ($\sim$10\textsuperscript{-2}\%) in the valleys, in excellent agreement with the designed stack. After annealing at 950 \textdegree{}C, the peak-to-valley amplitudes of all three isotope curves decay slightly, and a discernible rounding appears at the steep interfaces --- a subtle evolution that marks the onset of thermally activated solid-state bulk interdiffusion, in a temperature window where the vacancy and self-interstitial mechanisms make comparable contributions to silicon self-diffusion [\textcolor{blue}{16,17}]. This compositional profile shows that within the 850--950 \textdegree{}C window the two-dimensional layered geometry remains continuous and intact. The 850--950 \textdegree{}C interval therefore provides an effective and highly stable structural framework ideally suited to preserving the engineered isotope profile.

At higher temperatures, the structure becomes strongly homogenized. At 1050 \textdegree{}C, the sharp square wave is broadly flattened into a severely attenuated, low-amplitude profile; the valley positions of the minor isotopes (\textsuperscript{29}Si and \textsuperscript{30}Si) rise by several orders of magnitude, indicating that long-range bulk diffusion has driven vigorous intermixing between the ultra-thin 10-nm layers. This extensive chemical mixing --- the elimination of the initially steep isotopic concentration gradients and the complete depletion of the native oxide --- releases the internal micro-strain field, but evidently at the cost of interlayer confinement. When the temperature is further raised to the extreme condition of 1150 \textdegree{}C for 30 s, the periodic oscillations vanish completely; all three isotope curves broaden into absolutely flat horizontal lines converging precisely on their natural-abundance baselines (\textsuperscript{28}Si $\approx$ 92.2\%, \textsuperscript{29}Si $\approx$ 4.7\%, \textsuperscript{30}Si $\approx$ 3.1\%). This total annihilation of the multilayer cannot be attributed to conventional Fickian bulk diffusion alone. Rather, the 1150 \textdegree{}C thermal budget triggers morphological disintegration through Rayleigh capillary instability [\textcolor{blue}{18}], causing the continuous 10 nm films to rupture and dewet into isolated 3D hemispherical islands. At this extreme temperature, the liquid-like surface self-diffusivity of silicon adatoms, combined with rapid bulk mixing within the decoherent 3D micro-regions, jointly drive instantaneous compositional homogenization within the analysis depth. The AFM and SIMS results establish the thermal stability of the silicon isotope heterostructures: whereas the system maintains excellent planar and compositional integrity below 950 \textdegree{}C, higher temperatures trigger defect-mediated oxide rupture, structural disintegration, and complete macroscopic mixing.

\subsection{Low-temperature magnetotransport}

To evaluate the low-temperature electrical transport properties of the material, standard Hall-bar devices with a channel dimension of 100 \textmu{}m $\times$ 300 \textmu{}m (W $\times$ L) were fabricated on the enriched \textsuperscript{28}Si heterostructure. Phosphorus was implanted at a dose of 2 $\times$ 10\textsuperscript{15} cm\textsuperscript{-2} through a thermally grown 10 nm SiO\textsubscript{2} cap. Ohmic contact regions were defined by buffered-oxide etch (BOE), and Ti/Au metallization was deposited by electron-beam evaporation. A 30 nm Al\textsubscript{2}O\textsubscript{3} gate dielectric was grown by atomic-layer deposition (ALD) at 300 \textdegree{}C, with a patterned Al top gate defining the conductive channel. Importantly, this high-purity isotopic material platform has previously enabled an ultrafast coherent qubit operation in a collaborating group, underscoring its core strategic value for high-fidelity spin-qubit architectures.

\begin{figure*}[t]
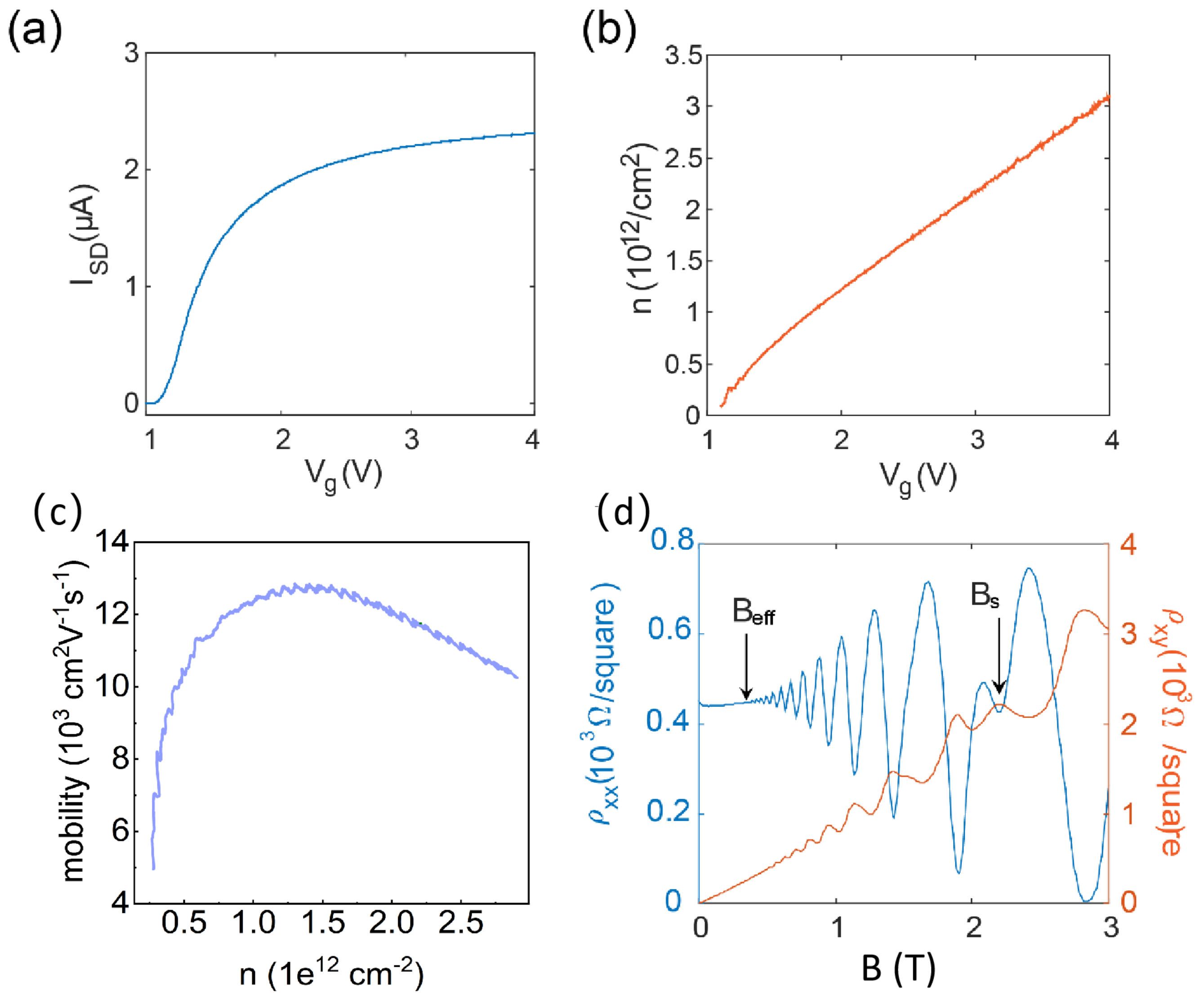

  \centering
  \FIG{5}{width=\textwidth}
  \caption{\textbf{Low-temperature magnetotransport and quantum-lifetime analysis of the enriched \textsuperscript{28}Si-MOS Hall-bar device.} (a) Source--drain current I\textsubscript{SD} versus top-gate voltage V\textsubscript{g}, showing a steep turn-on at V\textsubscript{th} $\approx$ 1.2 V and approaches saturation by 4 V. (b) Linear modulation of the 2DEG carrier density n with V\textsubscript{g}. (c) Hall mobility \textmu{} versus Hall carrier density n, peaking at $\sim$12,858 cm\textsuperscript{2} V\textsuperscript{-1} s\textsuperscript{-1} near n $\approx$ 1.3 $\times$ 10\textsuperscript{12} cm\textsuperscript{-2}. (d) Shubnikov--de Haas (SdH) oscillations of the longitudinal resistivity $\rho$\textsubscript{xx} (blue) and Hall resistivity $\rho$\textsubscript{xy} (orange) versus perpendicular field B (1--3 T) at V\textsubscript{g} = 2.5 V.}
  \label{5}
\end{figure*}

Figure~\ref{5}(a--d) shows the fundamental field-effect transport characteristics at temperature of 10 mK to 2700 mK. The device exhibits a clear turn-on threshold at V\textsubscript{th} $\approx$ 1.2 V, above which the source--drain current I\textsubscript{SD} rises rapidly and saturates at higher gate voltage (Fig.~\ref{5}(a)). The top-gate voltage linearly modulates the induced 2DEG carrier density n, extending smoothly from turn-on up to 3 $\times$ 10\textsuperscript{12} cm\textsuperscript{-2} (Fig.~\ref{5}(b)). The extracted Hall mobility \textmu{} evolves non-monotonically with carrier density n (Fig.~\ref{5}(c)). The mobility peaks at $\sim$12,858 cm\textsuperscript{2}V\textsuperscript{-1}s\textsuperscript{-1} near n $\approx$ 1.3 $\times$ 10\textsuperscript{12} cm\textsuperscript{-2}. The mobility approximately follows the power-law $\mu \propto $ with $\alpha $ from +0.87 to -0.45, indicating that transport is dominated by multiple scattering mechanisms, and the scattering competition mechanism changes with the variation of carrier density. For specific details, please refer to the power-law index piecewise fitting and analysis in the supplementary information section. To extract the deeper quantum-mechanical parameters of the induced \textsuperscript{28}Si-MOS channel, high-field perpendicular magnetotransport was analyzed. As shown in Fig.~\ref{5}(d), the longitudinal resistivity component exhibits robust and clearly resolved Shubnikov--de Haas (SdH) quantum oscillations over a 1--3 T field range[\textcolor{blue}{19}]. These oscillations onset at the very low field B\textsubscript{eff} $\approx$ 0.5 T (Fig.~\ref{5}(d)), confirming the low disorder within the enriched channel. In the high-field regime, $\rho$\textsubscript{xx} shows near-zero deep minima accompanied by corresponding Hall plateaus in $\rho$\textsubscript{xy}. Fitting the temperature-dependent damping of the SdH oscillation amplitude using the standard Lifshitz--Kosevich (LK) formula [\textcolor{blue}{20}] yields an electron effective mass of m* = 0.1872 m\textsubscript{0}, in excellent agreement with the transverse effective mass of bulk single-crystal silicon. Furthermore, Dingle analysis yields a quantum scattering time $\tau$\textsubscript{q} = 0.659 ps and a transport relaxation time $\tau$\textsubscript{t} = 1.48 ps (derived from the Hall mobility). The resulting single-particle level broadening, $\Gamma$ = 498 \textmu{}eV, is in excellent agreement with the classically reported value (480 \textmu{}eV) [\textcolor{blue}{21}]. In summary, these excellent magnetotransport metrics --- low turn-on voltage, minimal disorder-induced level broadening, and outstanding low-temperature mobility --- strongly demonstrate that the RPCVD-grown enriched \textsuperscript{28}Si film provides a promising material platform for constructing high-coherence spin-qubit devices.

\subsection{Spin-qubit demonstration}

\begin{figure*}[t]
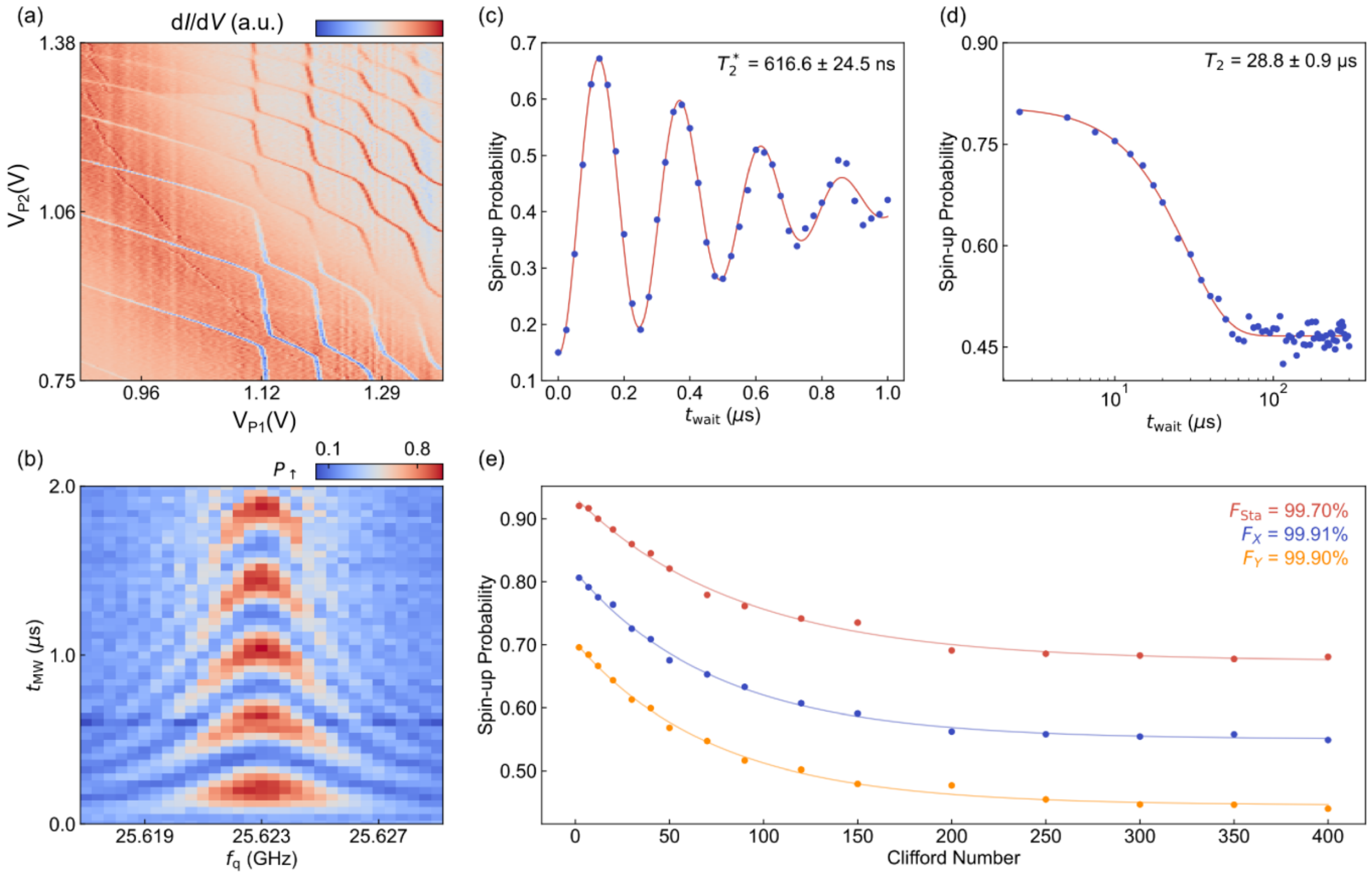

  \centering
  \FIG{6}{width=\textwidth}
  \caption{\textbf{Double-quantum-dot and single-qubit characterization of the enriched \textsuperscript{28}Si-MOS platform.} (a) Charge-stability diagram of the double quantum dot: d(SET)/dU\textsubscript{10} as a function of U\textsubscript{10} and U\textsubscript{8} (U\textsubscript{12} = 0.6125 V), showing the characteristic honeycomb pattern. (b) Rabi chevron: spin-up probability P$\uparrow$ versus qubit frequency and microwave pulse duration (0--2 \textmu{}s) at 7 dBm. (c) Ramsey free-induction decay, from which a dephasing time T\textsubscript{2}* = 616.6 $\pm$ 24.5 ns is extracted by fitting A\textperiodcentered{}exp(-(t/T)\textsuperscript{B})\textperiodcentered{}cos(2$\pi$ft)+P (stretch exponent B = 1.8). (d) Hahn-echo decay, yielding a coherence time T\textsubscript{2}\textsuperscript{Hahn} = 28.8 $\pm$ 0.9 \textmu{}s, from A\textperiodcentered{}exp(-(t/T)\textsuperscript{B})+P (stretch exponent B = 1.7). (e) Randomized benchmarking of the single-qubit Clifford gates: standard reference (99.70\%) and interleaved X (99.91\%) and Y (99.90\%) gates, plotted as spin-up probability versus Clifford number with exponential fits.}
  \label{6}
\end{figure*}

To demonstrate the quantum functionality of the platform, a lateral double quantum dot was fabricated on the enriched \textsuperscript{28}Si-MOS structure [\textcolor{blue}{40}]. Figure~\ref{6}(a) shows the charge-stability diagram of the double quantum dot, with the derivative of the single-electron transistor (SET) signal dI/dV plotted as a function of the gate voltages V\textsubscript{P1} and V\textsubscript{P2} (U\textsubscript{12} = 0.6125 V fixed), revealing the characteristic honeycomb pattern of well-formed double-dot charge configurations.

Coherent spin manipulation was performed at $\sim$20 mK using microwave excitation at 7 dBm (source amplitude). The Rabi chevron (Fig.~\ref{6}(b)) maps the spin-up probability P$\uparrow$ as a function of qubit frequency and microwave pulse duration (0--2 \textmu{}s), displaying the characteristic chevron pattern that confirms coherent Rabi oscillations and locates the qubit resonance frequency [\textcolor{blue}{6,32}]. Ramsey free-induction decay (Fig.~\ref{6}(c)) reveals damped oscillations from which a dephasing time \textbf{T\textsubscript{2}* = 616.6 $\pm$ 24.5 ns} is extracted by fitting a stretched-exponential envelope with a damped cosine. A Hahn-echo measurement (Fig.~\ref{6}(d)) suppresses quasi-static noise and yields a coherence time \textbf{T\textsubscript{2}\textsuperscript{Hahn} = 28.8 $\pm$ 0.9 \textmu{}s}, a more than forty-fold enhancement over T\textsubscript{2}*, consistent with the efficient refocusing of slowly fluctuating nuclear-spin and charge noise. Each data point was averaged over 20 repetitions of 300 single-shots, and the quoted uncertainties are the standard errors (1$\sigma$) of the individual fits.

Finally, single-qubit gate fidelity was quantified by randomized benchmarking [\textcolor{blue}{24,25}] (Fig.~\ref{6}(e)). Fitting the Clifford-number-dependent decay of the spin-up probability yields a standard (reference) fidelity of 99.70\%, and interleaved benchmarking of the X and Y ($\pi$) gates yields single-qubit average gate fidelities of 99.91\% and 99.90\%, respectively (i.e., up to $\sim$99.9\%; 95\% confidence interval); the interleaved gate fidelities are obtained from the decay-parameter ratio F = 1 - (1 - p\textsubscript{inter}/p\textsubscript{ref})/2, with p from P(m) = A\textperiodcentered{}p\textsuperscript{m} + B. The RB decay curves are vertically offset for clarity (reference +0.12, Y -0.12). Together, these results demonstrate single-qubit operation and establish RPCVD-grown \textsuperscript{28}Si as a promising material platform for silicon spin qubits [\textcolor{blue}{26--30}].

\section{Discussion}

We have demonstrated that an industrial RPCVD system can grow \textsuperscript{28}Si epilayers with an isotopic purity of 99.997\% (\textsuperscript{29}Si $\leq$ 30 ppm) --- approaching the level of specialized laboratory enrichment [\textcolor{blue}{7,8,10}] --- directly on high-resistivity FZ-Si substrates. The atomically flat, strain-free, defect-free interfaces revealed by AFM and HRTEM confirm that isotopic engineering at this purity does not compromise crystalline perfection, a prerequisite for low-disorder quantum transport [\textcolor{blue}{19}].

Our systematic annealing study establishes a clear thermal-budget window (below $\sim$950 \textdegree{}C) within which the isotope superlattice remains compositionally and structurally intact, providing a stable framework for preserving the engineered isotope profile. Above this window, we resolve a coherent physical picture in which defect-mediated native-oxide decomposition (Si + SiO\textsubscript{x} $\to$ 2SiO$\uparrow$) and vacancy injection first create localized voids, then --- upon oxide exhaustion --- give way to Rayleigh-instability-driven dewetting and complete intermixing. This mechanistic map is directly relevant to the thermal processing required to convert the epilayer into working devices.

The transport and qubit results close the loop. A peak Hall mobility of $\sim$12,858 cm\textsuperscript{2} V\textsuperscript{-1} s\textsuperscript{-1} and a quantum lifetime of 0.659 ps (level broadening 498 \textmu{}eV) place the induced 2DEG among the cleanest reported for MOS-gated silicon [\textcolor{blue}{21}]. More importantly, a double quantum dot on this platform already achieves T\textsubscript{2}* = 0.62 \textmu{}s, T\textsubscript{2}\textsuperscript{Hahn} = 28.8 \textmu{}s, and single-qubit gate fidelities of 99.9\% [\textcolor{blue}{6,26,33,34}]. These figures are competitive with early single-qubit demonstrations on laboratory-grown \textsuperscript{28}Si [\textcolor{blue}{7,26}] and, critically, they are obtained on an industrial, wafer-scale-capable platform.

Industrial-scale isotopic enrichment and CMOS-compatible RPCVD growth make this platform compatible with standard semiconductor manufacturing [\textcolor{blue}{13,22,31,37}], although wafer-level uniformity, thickness control, and integration with 300-mm CMOS substrates remain to be demonstrated. Further reductions in \textsuperscript{29}Si concentration, optimized gate-dielectric interfaces, and isotopic purification of the surrounding oxide will push coherence times and gate fidelities toward the fault-tolerance threshold [\textcolor{blue}{28,29,35,36,38}], and that extension of the demonstrated single-qubit control to two-qubit logic gates will position industrial RPCVD-grown \textsuperscript{28}Si as a cornerstone material for scalable silicon quantum processors.

\section{Methods}

\textbf{Epitaxial growth.} Enriched \textsuperscript{28}Si films were grown in an industrial reduced-pressure chemical vapour deposition (RPCVD) reactor at 650 \textdegree{}C on high-resistivity float-zone silicon (FZ-Si, resistivity $>$ 10,000 $\Omega$\textperiodcentered{}cm). The precursor, \textsuperscript{28}SiH\textsubscript{4}, was synthesized from gas-centrifuge-enriched SiF\textsubscript{4}. Two structures were grown: an $\sim$83-nm single \textsuperscript{28}Si homoepitaxial layer, and a \textsuperscript{28}Si/\textsuperscript{nat}Si isotope superlattice consisting of three 10-nm \textsuperscript{28}Si layers alternating with two 10-nm natural-silicon (natSi) barrier layers.

\section*{Acknowledgements}
This work was supported by the National Natural Science Foundation of China (Grant No. 62404248), the Outstanding Engineer Growth Program and the Innovation Program for Quantum Science and Technology (Grant No. 2021ZD0302300).


\end{document}